\documentclass[conference]{IEEEtran}
\IEEEoverridecommandlockouts
\usepackage{cite}
\usepackage{amsmath,amssymb,amsfonts}
\usepackage{algorithmic}
\usepackage{graphicx}
\usepackage{textcomp}
\usepackage{xcolor}
\usepackage{amsmath}
\usepackage{algorithm}
\usepackage{algorithmic}
\usepackage{svg}

\def\BibTeX{{\rm B\kern-.05em{\sc i\kern-.025em b}\kern-.08em
    T\kern-.1667em\lower.7ex\hbox{E}\kern-.125emX}}

\allowdisplaybreaks

\begin{document}

\title{GLRT for Reconfigurable Intelligent Surface aided Spectrum Sensing }
\author{Nikhilsingh Parihar,  Praful D. Mankar, and Sachin Chaudhari
\thanks{The authors are with Signal Processing and Communication Research Center, IIIT Hyderabad, India. Email: nikhilsingh.parihar@research.iiit.ac.in, \{praful.mankar,s.chaudhari\}@iiit.ac.in.} 
\vspace*{-1.2cm}
}

\maketitle
\begin{abstract}
Spectrum sensing (SS) is crucial for realising cognitive radio networks, where the secondary user (SU) needs to detect the presence of a primary user (PU) in order to utilise the spectrum. However, the ability of detection is influenced by unknown propagation environment factors such as multipath fading, correlated noise, transmission power of PU, etc. This paper investigates reconfigurable intelligent surfaces (RIS)-aided SS under correlated noise conditions using a generalised likelihood ratio test (GLRT) and energy detector (ED) frameworks. We first derive maximum likelihood estimates of the unknown channel state and transmit power, and employ these estimates to construct the GLRT-based test statistic using the signal received with an optimally configured RIS. The RIS phase shift matrix is optimally determined to maximise the gain of the estimated channel. 
Besides, the detection and false alarm probabilities of ED with optimally configured RIS are also derived. The numerical receiver output characteristics (ROC) demonstrate that the proposed GLRT achieves superior detection probability compared to ED, particularly under correlated noise and limited number of observations.

\end{abstract}
\begin{IEEEkeywords}
Spectrum sensing, reconfigurable intelligent surfaces, generalised likelihood ratio test, energy detector, cognitive radio.
\end{IEEEkeywords}

\section{Introduction}
\vspace{-2mm}
\label{sec:intro}
Reconfigurable Intelligent Surfaces (RISs) are expected to play a crucial role in the deployment of 6G networks \cite{Zhang_2022} because of their ability to dynamically control the wireless propagation environment. While RISs are primarily being explored for enhancing communication performance, their potential can also be leveraged for wireless sensing applications such as localisation and mapping \cite{wymeersch2020radio}, target detection \cite{buzzi2021radar}, and spectrum sensing \cite{xie2024enhancing}. This paper focuses on utilising RIS for spectrum sensing, which plays a key role in enabling cognitive radio for efficient spectrum utilisation, wherein the secondary user (SU) must detect the presence of a primary user (PU). Essentially, the RIS can improve the received signal strength by configuring its reflective elements to beamform the PU’s signal toward the SU, thereby enhancing detection reliability and sensing accuracy.

\textit{Related Works:} 
Since the formalization of cognitive radio, several detection methods have been widely studied in the literature for spectrum sensing including energy detection (ED) \cite{ma2009matchedfilter}, matched filtering (MF), cyclostationary-based detection \cite{jang2014cyclostationary}, and eigenvalue-based detectors \cite{bouallegue2018eigenvalues,zeng2008med}. The performance of these detectors fundamentally depends on the PU signal power and the propagation environment relative to the noise level. However, their performance can degrade in low SNR conditions, under noise uncertainty, with imperfect channel state information (CSI), or when there is no direct link between the PU and SU. To address scenarios involving unknown parameters, blind detection approaches such as the maximum eigenvalue detector (MED) and generalised likelihood ratio test (GLRT)-based methods have been explored \cite{lin2020glrt}. For example, \cite{patel2016glrt} investigates GLRT under uncertain CSI conditions and demonstrates improved detection, though it assumes uncorrelated noise for system modeling. In addition, in non-line-of-sight (NLoS) or obstructed environments, the received signal power at the SU can be significantly attenuated, resulting in reduced detection probability even at high SNR levels.

In such situations, RIS can substantially enhance spectrum sensing by introducing an additional controllable reflection path, effectively strengthening the received signal. Prior studies have shown that RIS-assisted systems outperform conventional setups without RIS. For instance, \cite{ge2022ris} employs the MED framework to analyse the number of RIS elements required to achieve a detection probability close to unity by leveraging optimal phase configurations. Similarly, works such as \cite{ge2023ris,xie2024enhancing} extend this analysis to active RIS architectures, where both phase adjustment and signal amplification are used to improve detection. However, these studies typically assume white Gaussian noise, perfect or statistical CSI, and rely on asymptotic test statistics, resulting in longer sensing durations and impractical channel assumptions. Although \cite{parihar2024rismed} addresses the correlated-noise scenario and provides analytical expressions that reduce dependence on asymptotic approximations and long sensing times. But, it assumes the PU–RIS channel to be perfectly known, which is impractical in realistic environments where the PU location or its channel parameters are typically unknown.
Although existing literature has independently investigated RIS-assisted sensing and GLRT-based detection, the combined use of GLRT in an RIS-aided environment under correlated noise and unknown channel conditions remains unexplored. To address these issues, this paper proposes a comprehensive GLRT-based spectrum sensing framework for RIS-assisted systems operating under correlated noise. The proposed approach jointly estimates the PU–RIS channel and the unknown PU transmit power through a maximum likelihood (ML) estimation procedure. The estimated parameters are then utilised to optimise the RIS phase shift matrix such that the effective channel gain at the SU is maximised. This joint estimation and RIS optimisation strategy enables robust and efficient spectrum sensing even when the PU–RIS channel and transmit power are unknown, providing improved detection performance under practical correlated noise environments.

\vspace{-1mm}
\section{System Model}
\vspace{-1mm}
This paper investigates a RIS-assisted wireless sensing system comprising a single-antenna PU, an $N$-antenna SU, and an $L$-element RIS. We employ a beyond-diagonal (BD) architecture for the RIS with interconnected elements to induce coupling, allowing for more flexible reflected signal control and enhanced channel strength. The BD-RIS configuration is defined by an $L \times L$ matrix $\boldsymbol{\Phi}$, where $\boldsymbol{\Phi}\boldsymbol{\Phi}^H = \mathbf{I}$.

The PU-RIS and RIS-SU channels are denoted by $\mathbf{h} \in \mathbb{C}^{L \times 1}$ and $\mathbf{G} \in \mathbb{C}^{N \times L}$, respectively. Considering a slow flat-fading environment, the channels are assumed to be constant over the sensing interval.
We assume the RIS to be an integral component of the spectrum sensing and it can assist the SU. Thus, it is safe to consider the RIS-SU channel $\mathbf{G}$ has dominant line-of-sight (LoS) component because optimal placement and it is known at the SU. Further, it is assumed that SU have control over the RIS phase-shift matrix $\mathbf{\Phi}$ to enable parameter estimation and spectrum sensing.
This paper focuses on developing the GLRT based solution wherein maximum likelihood estimates of unknown PU-RIS channel $\mathbf{h}$ response is required to perform the likelihood ratio test. However,  there is a fundamental limitation in estimation of $\mathbf{h}$ arising because of large dimensionality of RIS. Basically, to estimate $\mathbf{h}$ from the observed composite channel response $\mathbf{G\Phi h}$, it is essential to have the number of RIS elements $L$ lesser than or equal to the number of receiving antennas $N$. This is an impractical constraint as the RISs  are usually  larger in size. To overcome this issues, we employ a group-wise framework as discussed next.

In order to enable sensing with large RISs, the $L$ RIS elements are partitioned into $G$ disjoint groups, where each group contains 
$\bar{L} = L/G$ elements. 
 Let $\mathcal{I}_g$ denote the index set of RIS elements  corresponding to the $g$-th group, such that $\bigcup_{g=1}^{G}\mathcal{I}_g = \{1,2,\ldots,L\}$ and $\mathcal{I}_g \cap \mathcal{I}_{g'} = \emptyset$ for $g \neq g'$. During spectrum sensing, the RIS operates in a group-wise manner, sequentially, where only one RIS group is activated at a time while the remaining groups remain inactive. Hence, when the $g$-th group is active, the RIS phase-shift matrix is denoted by $\boldsymbol{\Phi}_g$, having non-zero entries only over the indices corresponding to $\mathcal{I}_g$.


The received signal at the SU during the $t$-th observation, when group $g$ is active, can be expressed as
\begin{equation}
\mathbf{y}_{g,t} = \sqrt{\nu}\mathbf{G}_g\boldsymbol{\Phi}_g\mathbf{h}_g x_t + \mathbf{n}_t,
\qquad t = 1,2,\ldots,T_g,
\label{eq:received_signal_group}
\end{equation}
where $\nu = (d_1 d_2)^{-\xi}$ represents path-loss attenuation along the PU-RIS-SU link,  $d_1$ and $d_2$ are the PU-RIS and RIS-SU distances, respectively, $\xi$ represents the path-loss exponent, $\mathbf{h}_g$ and $\mathbf{G}_g$ denote the PU-RIS sub-channel vector and RIS-SU   sub-channel matrix corresponding to the $g$-th group, respectively. Next, $x_t \sim \mathcal{CN}(0,\sigma_x^2)$ denotes the transmitted symbol where the transmit power $\sigma_x^2$ is unknown and $\mathbf{n}_t \sim \mathcal{CN}(\mathbf{0},\mathbf{R}_n)$ denotes complex correlated gaussian noise where $\mathbf{R}_n$ is the noise covariance matrix.

The proposed RIS-assisted spectrum sensing method operates in the following two phase framework. 
\begin{enumerate}
    \item In the first phase, the SU performs channel estimation and RIS configuration, wherein the PU--RIS sub-channel vectors $\mathbf{h}_g$ and the PU transmit power $\sigma_x^2$ are estimated and utilized to optimize RIS phase-shift matrices $\boldsymbol{\Phi}$ for maximizing the effective received signal power at SU.
    \item In the second phase, the optimally configured RIS is used for spectrum sensing, wherein the SU collects group-wise observations under optimally configured RIS and performs hypothesis testing using GLRT and ED.
\end{enumerate}

\begin{figure}[htbp]
  \centering\vspace{-3mm}
  \includegraphics[width=0.25\textwidth]{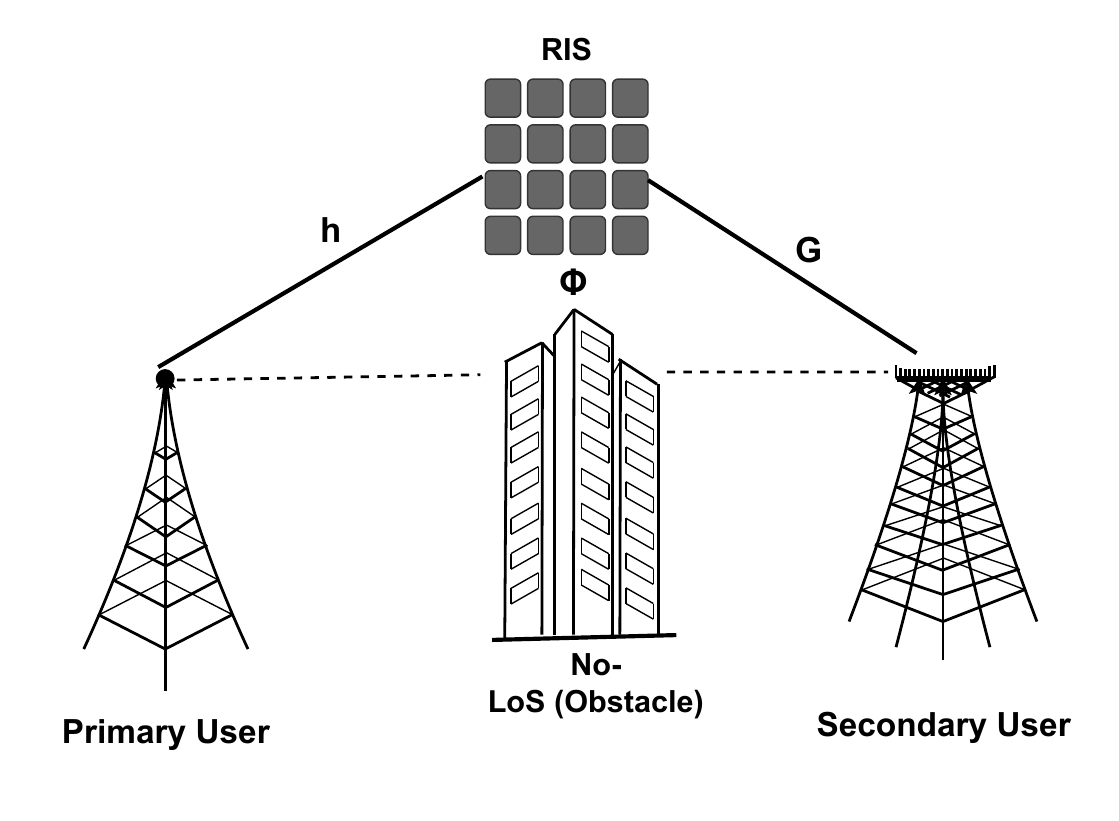}\vspace{-3mm}
  \caption{System model of the RIS-assisted spectrum sensing framework.}
  \label{fig:sys_model}
\end{figure}

The hypothesis testing problem for the spectrum sensing under $g$-th  group can be formulated as
\begin{subequations}
\begin{align}
\mathcal{H}_0 &:~ \mathbf{y}_{g,t} = \mathbf{n}_t, \label{eq:Ho_g}\\
\mathcal{H}_1 &:~ \mathbf{y}_{g,t} = \sqrt{\nu}\mathbf{G}_g\boldsymbol{\Phi}_g\mathbf{h}_g x_t + \mathbf{n}_t. \label{eq:H1_g}
\end{align}
\end{subequations}
The  sample covariance matrix of $\mathbf{y}_{g,t}$ associated with $T_g$ observations can be determined as
\begin{equation}
\hat{\mathbf{R}}_{y,g} = \frac{1}{T_g}\sum\nolimits_{t=1}^{T_g}\mathbf{y}_{g,t}\mathbf{y}_{g,t}^H.
\end{equation}
For given $\mathbf{h}_g$ and $\sigma_x^2$, the exact covariance matrix under $\mathcal{H}_1$ is given by
\begin{equation}
\boldsymbol{\Sigma}_{y,g}(\mathbf{h}_g,\sigma_x^2)
= \sigma_x^2 \mathbf{G}_g\boldsymbol{\Phi}_g\mathbf{h}_g\mathbf{h}_g^H\boldsymbol{\Phi}_g^H\mathbf{G}_g^H + \mathbf{R}_n.
\end{equation}
It is important to note that the covariance matrix $\boldsymbol{\Sigma}_{y,g}(\mathbf{h}_g,\sigma_x^2)$ depends jointly on the unknown PU-RIS sub-channel vector $\mathbf{h}_g$ and the unknown transmit power $\sigma_x^2$. This group-wise signal model forms the basis for the parameter estimation and detection system which will be discussed in the next section.

\section{Parameter Estimation}
\label{sec:estimation}
\vspace{-2mm}
For spectrum sensing, we first obtain a joint estimate of the PU-RIS sub-channel vectors $\mathbf{h}_g$ and the PU transmit power $\sigma_x^2$ based on the received signal which are  observed at the SU. Then the estimate for each sub vector are stacked to obtain complete channel vector $\hat{\mathbf{h}}$, need to change. 
We first whitened the the received signal and then apply  maximum likelihood (ML) estimation.
The  observations received under the $g$-th RIS group are prewhiten as
\begin{equation}
    \tilde{\mathbf{y}}_{g,t}
    = \mathbf{W}\mathbf{y}_{g,t}
    = \sqrt{\nu}\mathbf{A}_g\mathbf{h}_g x_t + \tilde{\mathbf{n}}_t,
    \label{eq:whitened_group}
\end{equation}
where $\mathbf{W} = \mathbf{R}_n^{-\frac{1}{2}}$,
$\mathbf{A}_g = \mathbf{W}\mathbf{G}_g\boldsymbol{\Phi}_g$,
and $\tilde{\mathbf{n}}_t = \mathbf{W}\mathbf{n}_t$ so that
$\tilde{\mathbf{n}}_t \sim \mathcal{CN}(\mathbf{0},\mathbf{I})$.
The corresponding sample covariance matrix for $T_g$ whitened observations is given by
\begin{equation}
    \hat{\mathbf{R}}_{\tilde{y},g}
    = \frac{1}{T_g}\sum\nolimits_{t=1}^{T_g}
    \tilde{\mathbf{y}}_{g,t}\tilde{\mathbf{y}}_{g,t}^H.\label{eq:cov_y_tiled}
\end{equation}
Hence, the covariance of $\tilde{\mathbf{y}}_{g,t}$ under $\mathcal{H}_1$ becomes
\begin{equation}
    \boldsymbol{\Sigma}_{\tilde{y},g}
    = \mathbb{E}[\tilde{\mathbf{y}}_{g,t}\tilde{\mathbf{y}}_{g,t}^H|\mathcal{H}_1]
    = \mathbf{I} + \sigma_{x}^2 \mathbf{A}_g\mathbf{h}_g\mathbf{h}_g^H\mathbf{A}_g^H.
    \label{eq:SigmaW_group}
\end{equation}
Clearly, the vectors $\tilde{\mathbf{y}}_{g,t}$ are independent and identically distributed complex Gaussian with covariance $\boldsymbol{\Sigma}_{\tilde{y},g}$ and thus, the log-likelihood function of parameters $\mathbf{h}_g$ and $\sigma_{x}^2$ for the $g$-th group can be written as
\begin{equation}
    \mathcal{L}_g(\mathbf{h}_g,\sigma_x^2)
    = -T_g \log|\boldsymbol{\Sigma}_{\tilde{y},g}|
      - T_g\,\mathrm{tr}\left(
      \boldsymbol{\Sigma}_{\tilde{y},g}^{-1}\hat{\mathbf{R}}_{\tilde{y},g}
      \right)
      + \text{c.}
    \label{eq:logL_group}
\end{equation}
Now by exploiting the structure of $\mathbf{\Sigma}_{\tilde{y}}$ given in \eqref{eq:SigmaW_group} and using standard matrix identities, we can simplify  maximize the above likelihood function to obtain the ML estimates. The resulting estimates are given below.
\begin{enumerate}
    \item The estimate of PU-RIS sub-channel vector $\mathbf{h}_{g}$ is
    \vspace{-2mm}
    \begin{equation}
        \hat{\mathbf{h}}_g
        = \arg\max_{\mathbf{h}_g}
        \frac{\mathbf{h}_g^H \mathbf{A}_g^H \hat{\mathbf{R}}_{\tilde{y},g} \mathbf{A}_g\mathbf{h}_g}
        {\mathbf{h}_g^H \mathbf{A}_g^H\mathbf{A}_g\mathbf{h}_g}.
        \label{eq:rayleigh_group}
    \end{equation}
    The solution to \eqref{eq:rayleigh_group} is given by the dominant generalized eigenvector of the matrix pair
    $(\mathbf{A}_g^H \hat{\mathbf{R}}_{\tilde{y},g}\mathbf{A}_g,
      \mathbf{A}_g^H\mathbf{A}_g)$.

    \item
    Given $\mathbf{h}_g=\hat{\mathbf{h}}_g$, the ML estimate of the PU transmit power is
    \begin{equation}
        \hat{\sigma}_{x,g}^2
        =
        \frac{\hat{\mathbf{h}}_g^H \mathbf{A}_g^H \hat{\mathbf{R}}_{\tilde{y},g} \mathbf{A}_g\hat{\mathbf{h}}_g
        - \hat{\mathbf{h}}_g^H \mathbf{A}_g^H\mathbf{A}_g\hat{\mathbf{h}}_g}
        {(\hat{\mathbf{h}}_g^H \mathbf{A}_g^H\mathbf{A}_g\hat{\mathbf{h}}_g)^2}.
        \label{eq:sigma_est_group}
    \end{equation}
\end{enumerate}
Please refer to Appendix~\ref{appendix:ML_derivation} for a detailed derivation of \eqref{eq:rayleigh_group} and \eqref{eq:sigma_est_group}. Note that the parameters $\mathbf{h}_g$ and $\sigma_{x}^2$ appear in the log-likelihood function only through the product $\sigma_{x}^2\mathbf{A}_g\mathbf{h}_g\mathbf{h}_g^H\mathbf{A}_g^H$, see \eqref{eq:appendix_glrt_cov_g}. The pair $(\alpha\mathbf{h}_g,\,\sigma_{x}^2/|\alpha|^2)$ produces the same likelihood value for any scalar
$\alpha\neq0$, leading to an indefinite joint estimation.
Hence, to ensure an unique solution, we will impose the following normalization constraint
\begin{equation}
    \mathbf{h}_g^H \mathbf{A}_g^H\mathbf{A}_g\mathbf{h}_g = 1.
    \label{eq:constraint_group}
\end{equation}
Under this constraint, the transmit power estimate in \eqref{eq:sigma_est_group} simplifies to
\vspace{-2mm}
\begin{equation}
    \hat{\sigma}_{x,g}^2 = \max(\lambda_{\max,g}-1,\,0),
    \label{eq:final_sigma_group}
\end{equation}
where $\lambda_{\max,g}$ denotes the largest generalized eigenvalue of the matrix pair
$(\mathbf{A}_g^H \hat{\mathbf{R}}_{\tilde{y},g}\mathbf{A}_g,
 \mathbf{A}_g^H\mathbf{A}_g)$.
This normalization ensures that the solution $\hat{\mathbf{h}}_g$ and $\hat{\sigma}_{x}^2$
are unique and invariant to any shift in scaling.
Finally, the estimate of PU transmission power  can be obtained as
\begin{equation}
     \hat{\sigma}_{x}^2=\frac{1}{G}\sum\nolimits_{g=1}^G  \hat{\sigma}_{x,g}^2,
\end{equation}
and the estimate of PU-RIS channel  can be obtained by stacking up $\hat{\mathbf{h}}_g$'s, i.e.
\begin{equation}
     \hat{\mathbf{h}}=\left[\hat{\mathbf{h}}_1^T,\dots,\hat{\mathbf{h}}_G^T\right]^T.\label{eq:h_hat}
\end{equation}

\section{SNR Maximization}
\label{sec:SNR_maximization}
\vspace{-2mm}
The effective channel observed at the SU can be expressed approximately as
$\tilde{\mathbf{G}} \boldsymbol{\Phi} \hat{\mathbf{h}}$,
where $\tilde{\mathbf{G}} = \mathbf{W}\mathbf{G}$. Thus, to enhance the received signal strength for improving sensing capability, it is important to maximize the gain of the estimated channel $\tilde{\mathbf{G}}\mathbf{\Phi}\hat{\mathbf{h}}$ by configuring the RIS phase-shift optimally.
Thus, the RIS optimization problem can be formulated as
\begin{subequations}
    \begin{align}
        \max_{\mathbf{\Phi}} \quad
        & \| \tilde{\mathbf{G}} \boldsymbol{\Phi} \hat{\mathbf{h}} \|^2,
        \label{eq:Phi_max_g} \\
        \text{s.t.} \quad
        & \boldsymbol{\Phi} \boldsymbol{\Phi}^{H} = \mathbf{I},
        \label{eq:phi_constraint_g}
    \end{align}
\end{subequations}
where \eqref{eq:phi_constraint_g} enforces the unitary constraint of the BD-RIS over the active group.
The objective function in \eqref{eq:Phi_max_g} can be written as
\begin{equation*}
    J
    = \| \tilde{\mathbf{G}} \mathbf{\Phi} \hat{\mathbf{h}} \|^2
    = \mathrm{Tr}\left(
    \tilde{\mathbf{G}} \boldsymbol{\Phi}
    \hat{\mathbf{h}} \hat{\mathbf{h}}^{H}
    \boldsymbol{\Phi}^{H} \tilde{\mathbf{G}}^{H}
    \right).
\end{equation*}
Following a similar approach to \cite{ManasaICC}, we obtain a closed-form solution for $\mathbf{\Phi}$ that maximizes $J$.
Using the eigenvalue decompositions $\tilde{\mathbf{G}}^{H}\tilde{\mathbf{G}} = \mathbf{U}_{\tilde{G}}\mathbf{\Sigma}_{\tilde{G}}\mathbf{U}_{\tilde{G}}^{H}$ and $\hat{\mathbf{h}}\hat{\mathbf{h}}^{H} = \mathbf{U}_{\rm{h}}\mathbf{\Sigma}_{\rm{h}}\mathbf{U}_{\rm{h}}^{H}$ and cyclic property of trace, we can write $J$ as
\begin{align*}
    J
    &= \mathrm{Tr}\left(
    \boldsymbol{\Sigma}_{\tilde{G}}
    \mathbf{U}_{\tilde{G}}^{H}
    \boldsymbol{\Phi}
    \mathbf{U}_{{\rm h}}
    \boldsymbol{\Sigma}_{{\rm h}}
    \mathbf{U}_{{\rm h}}^{H}
    \boldsymbol{\Phi}^{H}
    \mathbf{U}_{\tilde{G}}
    \right) \\
    & \leq \mathrm{Tr}\left(
    \boldsymbol{\Sigma}_{\tilde{G}}\boldsymbol{\Sigma}_{{\rm h}}
    \right),
\end{align*}
where the inequality follows from the Von Neumann trace inequality.
The upper bound is achieved, while satisfying \eqref{eq:phi_constraint_g} by choosing the RIS phase-shift matrix is given as
\begin{equation}
    \boldsymbol{\Phi}^{\star}
    = \mathbf{U}_{{\tilde{G}}}\mathbf{U}_{{\rm h}}^{H}.
    \label{eq:Phi_opt_g}
\end{equation}
Substituting $\boldsymbol{\Phi}^{\star}$ into the objective function gives
\begin{equation}
    J_{\max}
    = \mathrm{Tr}\left(
    \boldsymbol{\Sigma}_{\tilde{G}}\boldsymbol{\Sigma}_{{\rm h}}
    \right)
    \stackrel{(a)}{=}
    \lambda_{\tilde{G}}\lambda_{{\rm h}},
    \label{eq:Phi_final_g}
\end{equation}
where step (a) follows from the fact that
$\tilde{\mathbf{G}}^{H}\tilde{\mathbf{G}}$ and
$\hat{\mathbf{h}}\hat{\mathbf{h}}^{H}$ are rank-one matrices with
$\lambda_{\tilde{G}}$ and $\lambda_{{\rm h}}$ being their respective non-zero eigenvalues.

\vspace{-2mm}
\section{Signal Detection}
\vspace{-2mm}
With optimally configured RIS based on the estimated PU-RIS channel vectors, the SU observes the received signals and detects the presence of the primary signal.  The detection problem with noise whitening can be written as
\begin{subequations}\label{eq:hypo_re_g}
\begin{align}
    \mathcal{H}_0 &: \tilde{\mathbf{y}}_{t} = \tilde{\mathbf{n}}_t, \\
    \mathcal{H}_1 &: \tilde{\mathbf{y}}_{t}
    = \mathbf{A}^{\star}\mathbf{h} x_t + \tilde{\mathbf{n}}_t,
\end{align}
\end{subequations}
\noindent where
$\mathbf{A}^{\star} = \sqrt{\nu}\mathbf{W}\mathbf{G}\boldsymbol{\Phi}^{\star}$
denotes the effective sensing matrix corresponding to the optimally configured RIS.
We employ GLRT and ED  for the above binary hypothesis testing problem to study the tradeoff between detection performance and computational complexity.
\vspace{-1mm}
\subsection{Generalised Likelihood Ratio Test}
\vspace{-2mm}
Leveraging the estimates of the PU-RIS channel vectors and the PU transmit power, we adopt the GLRT for signal detection problem given in \eqref{eq:hypo_re_g}.
The GLRT test statistic is defined as the maximized log-likelihood ratio for $\mathcal{H}_1$ and $\mathcal{H}_0$, i.e.,
\begin{equation}
    {\rm TS}_{\mathrm{GLRT}}
    =
    \frac{
    \max_{\mathbf{h},\sigma_x^2}
    \, \mathbf{p}(\tilde{\mathbf{Y}};\mathbf{h},\sigma_x^2,\mathcal{H}_1)
    }{
    \mathbf{p}(\tilde{\mathbf{Y}};\mathcal{H}_0)
    },
    \label{eq:TS_GLRT_g}
\end{equation}
where $\tilde{\mathbf{Y}} = [\tilde{\mathbf{y}}_{1},\ldots,\tilde{\mathbf{y}}_{T}]$,  $\mathbf{p}(\tilde{\mathbf{Y}};\mathbf{h},\sigma_x^2,\mathcal{H}_1)$   is the likelihood of parameters $\mathbf{h}$ and $\sigma_x^2$  under $\mathcal{H}_1$ and $\mathbf{p}(\tilde{\mathbf{Y}};\mathcal{H}_0)$ is the likelihood function under $\mathcal{H}_0$.
Performing the maximization of the corresponding log-likelihood ratio in \eqref{eq:TS_GLRT_g} with respect to the unknown parameters $\mathbf{h}$ and $\sigma_x^2$ yields the  GLRT test statistic as
\begin{equation}
    {\rm TS}_{\mathrm{GLRT}}
    = T\big(\hat{\lambda}_{\max} - \ln\hat{\lambda}_{\max}\big),
    \label{eq:GLRT_statistic_g}
\end{equation}
$\hat{\lambda}_{\max}$ denotes the dominant generalized eigenvalue of the matrix pair
$(\mathbf{A}^{\star H}\hat{\mathbf{R}}_{\tilde{y}}\mathbf{A}^{\star},
 \mathbf{A}^{\star H}\mathbf{A}^{\star})$.
A detailed derivation of \eqref{eq:GLRT_statistic_g} is provided in Appendix~\ref{appendix:GLRT_derivation}.
Finally, the GLRT decision rule can be written as
\begin{equation}
    {\rm TS}_{\mathrm{GLRT}}
    \underset{\mathcal{H}_0}{\overset{\mathcal{H}_1}{\gtrless}}
    \gamma_{\mathrm{GLRT}},
    \label{eq:GLRT_test}
\end{equation}
where $\gamma_{\mathrm{GLRT}}$ is the detection threshold chosen to satisfy the desired false-alarm probability 
\[
P_{\mathrm{FA}}
= \mathbb{P}({\rm TS}_{\mathrm{GLRT}}>\gamma_{\mathrm{GLRT}};\mathcal{H}_0),
\] 
and the corresponding detection probability
\[
P_{\mathrm{D}}
= \mathbb{P}({\rm TS}_{\mathrm{GLRT}}>\gamma_{\mathrm{GLRT}};\mathcal{H}_1).
\]
Due to the difficulty in obtaining the exact distribution of ${\rm TS}_{\mathrm{GLRT}}$ under both hypotheses, deriving the closed-form expressions for $P_{\mathrm{FA}}$ and $P_{\mathrm{D}}$ becomes analytically intractable. Therefore, the detection performance of the proposed GLRT is evaluated numerically by plotting the receiver operating characteristic (ROC)  in Section~\ref{sec:results}.

\subsection{Energy Detector}
\vspace{-2mm}
Here, we present the ED for signal detection with optimally configured RIS. During sensing, the RIS is configured based on the estimated PU-RIS channel vectors $\hat{\mathbf{h}}$. Consequently, the received signal statistics under the optimized RIS configuration depend on the accuracy of channel estimation. The optimal RIS
phase-shift matrix $\boldsymbol{\Phi}^{\star}$ obtained in
\eqref{eq:Phi_opt_g} provides the maximized gain of the effective channel as
$\|\mathbf{A}^{\star}\mathbf{h}\|^2
= \lambda_{{\rm h}}\lambda_{{\tilde{G}}}$,
where $\lambda_{{\rm h}}$ denotes the non-zero eigenvalue of
$\mathbf{h}\mathbf{h}^H$ (see \eqref{eq:Phi_final_g}).
Since the channel is assumed to be slow fading, the effective channel response
$\mathbf{A}^{\star}\mathbf{h}$ remains constant over the observation period.
Thus, for a given $\mathbf{h}$, the distribution of the whitened received
signal under the two hypotheses can be expressed as
\begin{subequations}
\begin{align}
    \mathcal{H}_0 &:
    \tilde{\mathbf{y}}_{t} \sim \mathcal{CN}(\mathbf{0},\mathbf{I}),\\
    \mathcal{H}_1 &:~
    \tilde{\mathbf{y}}_{t} \sim
    \mathcal{CN}(\mathbf{0},\mathbf{R}_{\tilde{y}}),
\end{align}
\end{subequations}
where
$
\mathbf{R}_{\tilde{y}}
= \sigma_x^2
\mathbf{A}^{\star}\mathbf{h}\mathbf{h}^H\mathbf{A}^{\star H}
+ \mathbf{I}.
$
The ED statistic for the same can be obtained using \eqref{eq:cov_y_tiled} as
\begin{equation}
    {\rm TS}_{\mathrm{ED}}
    = \frac{1}{N}\mathrm{tr}(\hat{\mathbf{R}}_{\tilde{y}})
    = \frac{1}{NT}
    \sum_{t=1}^{T}
    \|\tilde{\mathbf{y}}_{t}\|^2.
\end{equation}
Since ${\rm TS}_{\mathrm{ED}}$ is a sum of $NT$ independent complex Gaussian random variables  (conditioned on $\mathbf{h}$), the central limit theorem (CLT) allows us to approximate
${\rm TS}_{\mathrm{ED}}$ as Gaussian under both hypotheses:
\begin{subequations}
\begin{align}
    \mathcal{H}_0 &:~ {\rm TS}_{\mathrm{ED}} \sim 
    \mathcal{N}\left(\mu_0,\, \sigma_0^2\right),  \\
    \mathcal{H}_1 &:~ {\rm TS}_{\mathrm{ED}} \sim 
    \mathcal{N}\left(\mu_1,\, \sigma_1^2;{\mathbf{h}}\right), 
\end{align}\label{eq:statistics_ED}
\end{subequations}
where
$\mu_0 = 1$,
$\sigma_0^2 = \tfrac{1}{NT}$,
$\mu_{1}(\lambda_{{\rm h}}) = \tfrac{1}{N}\mathrm{tr}(\mathbf{R}_{\tilde{y}})$,
and
$\sigma_{1}^2(\lambda_{{\rm h}})
= \tfrac{1}{N^2T}\mathrm{tr}(\mathbf{R}_{\tilde{y}}^2)$. Note that the matrix $\mathbf{A}^{\star}\mathbf{h}\mathbf{h}^H\mathbf{A}^{\star H}$ is rank-1 matrix since it is outer product of $\mathbf{A}^{\star}\mathbf{h}$, hence it has the maximum eigenvalue as $1 + \sigma_{x}^{2}||\mathbf{A}^{\star}\mathbf{h}||^{2}$ and $N-1$ remaining  are equal to 1. Further, exploiting this structure for eigenvalue, it is obvious that the mean and variance under $\mathcal{H}_1$ are function of $||\mathbf{A}^{\star}\mathbf{h}||^{2}$ and using \eqref{eq:Phi_final_g} shows that $\mu_{1}$ and $\sigma_{1}^2$ are functions of $\lambda_{h}$.
Thus, ED statistics can be approximated as Gaussian under both hypotheses as
\begin{subequations} 
\begin{align} 
\mathcal{H}_0 &:~ {\rm TS}_{\mathrm{ED}} \sim \mathcal{N}(\mu_0,\sigma_0^2),\\ 
\mathcal{H}_1 &:~ {\rm TS}_{\mathrm{ED}} \sim \mathcal{N}\left(\mu_{1}(\lambda_{{\rm h}}),\sigma_{1}^2(\lambda_{{\rm h}})\right), 
\end{align} 
\end{subequations}
Finally, the ED decision rule can be written as
\begin{equation}
    {\rm TS}_{\mathrm{ED}}
    \underset{\mathcal{H}_0}{\overset{\mathcal{H}_1}{\gtrless}}
    \gamma_{\mathrm{ED}},
\end{equation}
where the threshold $\gamma_{\mathrm{ED}}$ is selected to satisfy a desired
global false-alarm probability, which can be obtained using \eqref{eq:statistics_ED} as 
\begin{equation}
    P_{\mathrm{FA}}
    =
    Q\left(
    \frac{\gamma_{\mathrm{ED}} - \mu_0}
         {\sigma_0}
    \right).
\end{equation}
For a given $P_{\mathrm{FA}}$,  the  threshold can be obtained as
\begin{equation}
    \gamma_{\mathrm{ED}}
    =
    G\mu_0
    +
    \sigma_0\,
    Q^{-1}(P_{\mathrm{FA}}).
\end{equation}
Using \eqref{eq:statistics_ED}, the detection probability can be determined as
\begin{equation}
    P_{\mathrm{D}}(\gamma_{\mathrm{ED}})
    =
    \mathbb{E}_{\lambda_{{\rm h}}}
    \left[
    Q\left(
    \frac{\gamma_{\mathrm{ED}}
    - \mu_{1}(\lambda_{{\rm h}})}
    {\sigma_{1}(\lambda_{{\rm h}})}
    \right)
    \right],\label{eq:ED_pd}
\end{equation}
where the expectation accounts for the deconditioning of the PU--RIS channel
sub-vectors across different RIS groups. The distribution of $\lambda_{{\rm h}}$  given in \cite[Theorem~1]{parihar2024rismed} can be used to evaluate the expectation in \eqref{eq:ED_pd}.

\section{Results}
\label{sec:results}
\vspace{-2mm}
In this section, we present numerical analysis for the proposed BD-RIS assisted GLRT and ED methods for spectrum sensing. 
First, we present the mean-squared error (MSE) performance of proposed MLE  for PU-RIS channel $\mathbf{h}$ and PU transmission power $\sigma_x^2$. Next, we present the ROC analysis of the proposed detection methods. For the numerical analysis, we consider the number of received antenna $N = 8$, the number of RIS element $L = 512$, the number of partition groups $G = 16$, the pathloss exponent $\xi = 3$, noise correlation factor $\rho=0.3$ and noise variance $\sigma_n^2=5$; unless mentioned otherwise. The noise covariance matrix is modeled as 
$\mathbf{R_{n,}}_{ij} = \rho^{\left|i-j\right|}\sigma_{n}^{2}$.

\begin{figure}[htbp]
  \centering
    \includegraphics[width=0.36\textwidth]{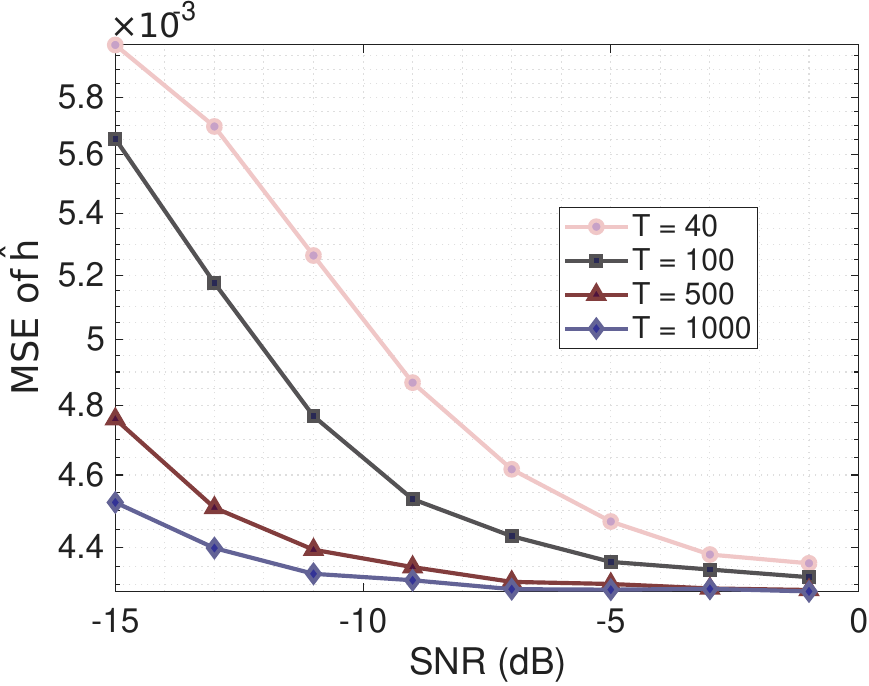}\vspace{-3mm}
  \caption{MSE of $\hat{\mathbf{h}}$ versus SNR for different numbers of observations $T$.}
  \label{fig:mmse_vs_snr}
\end{figure}

Fig. \ref{fig:mmse_vs_snr}  shows the MSE  performance of the proposed group-wise ML estimator for PU–RIS channel  estimate $\hat{\mathbf{h}}$ for different numbers of  observations $T \in \{40, 100, 500, 1000\}$. Here, the number of RIS elements  $L=512$ are grouped into $G=8$ groups of size $\bar{L}=N=64$. The graph shows that the MSE decreases monotonically with increasing SNR and increasing the number of observations  $T$, as expected. This verifies that the proposed grouped RIS architecture provides reliable channel estimation.

\begin{figure}[htbp]
  \centering
    \includegraphics[width=0.36\textwidth]{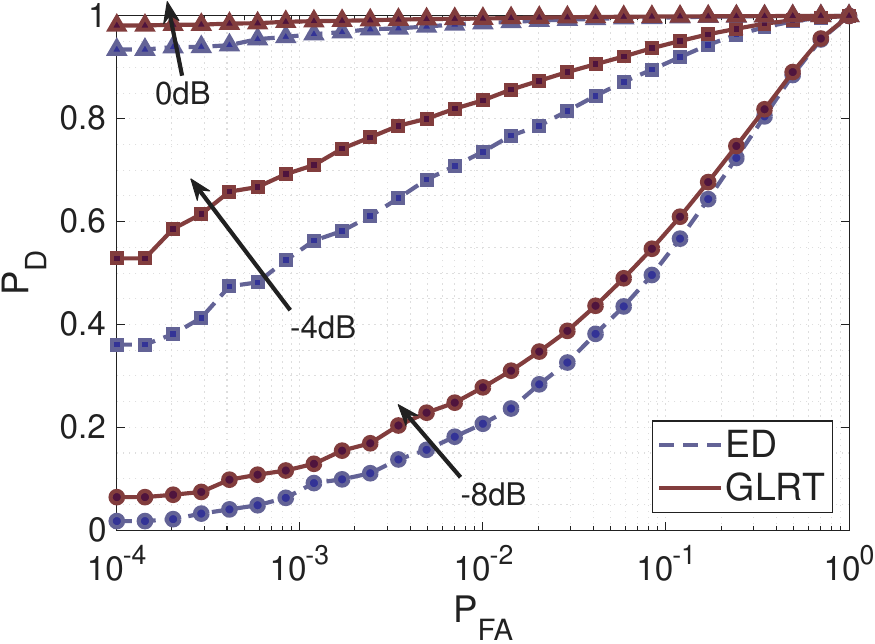}
  \caption{ROC curves for proposed GLRT and ED  for $N=8$ and $L=16$. The circular, square and triangular markers represent SNR equal to -8, -4 and 0 dB, respectively.}
   \label{fig:roc_curves}
\end{figure}

Fig. \ref{fig:roc_curves} shows the ROC curves of the proposed GLRT and ED detectors for  SNR equal to $0, -4$ and $-8$ dB. 
It can be observed from the figure that  the GLRT consistently outperforms the ED for all SNRs. This is because the GLRT exploits the estimated channel state and PU transmission power information to accurately characterize the likelihood function with the optimally configured RIS, whereas ED uses only information of the energy of the samples  received under optimized RIS. With the increasing SNR values, the detection performance of both detectors improves, while GLRT maintains a clear advantage in terms of detection probability for a given false-alarm rate. Further, it can be seen that the detection probability increases gradually with the increase in false alarm probability at low SNR. This is because at low SNR the signal strength is weak and hence the received signal in both hypotheses are quite similar, making it difficult to detect the presence of the signal. 

\begin{figure}[htbp]
  \centering
    \includegraphics[width=0.4\textwidth]{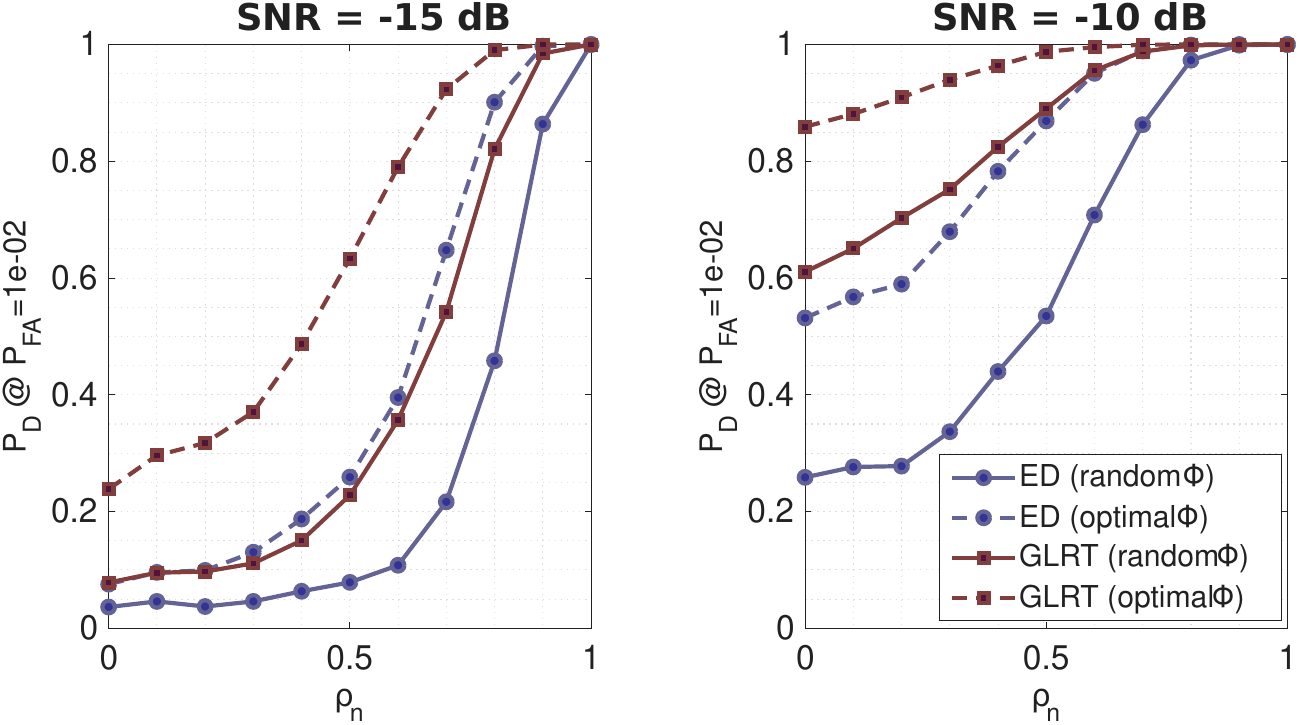}
  \caption{Probability of Detection versus the noise correlation coefficient.}
  \label{fig:rho_vs_pd}
\end{figure}

Finally,  we analyze the impact of noise correlation $\rho_n$ on the detection performance in Fig.~\ref{fig:rho_vs_pd} for the false-alarm probability $P_{FA}=10^{-2}$ and SNR equal to $-15$ and $-10$~dB. For each SNR, GLRT and ED detectors are assessed under random and optimal RIS phase configurations, in order to understand the impact of optimal RIS configuration.
As expected, for both the values SNR, $P_D$ increases monotonically with $\rho_n$ due to stronger noise correlation, which eventually helps to  effectively reduces the noise spread and thus improves the distinction between noise-only and signal-present hypotheses post whitening. Clearly, the higher SNR values further enhances detection performance across all schemes.
Further, the figure also shows that the optimal RIS phase configuration performs better than the random phase configuration for any $\rho_n$ and for both detectors. This is again attributed to the fact that the optimally configured RIS facilitates signal reception with better strength. The GLRT with optimal RIS phases achieves the highest $P_D$ by leveraging the channel structure knowledge along with the increase effective channel gain. It is important to note that the GLRT with random RIS phases still outperforms ED with optimal RIS configuration. In overall, these results confirm that the proposed grouped BD-RIS framework with GLRT and optimal phase design offers a significant performance gains under practical conditions.

\vspace{-4mm}
\section{Conclusion}
\vspace{-2mm}
This paper studied spectrum sensing in a RIS-aided cognitive radio system with a grouped beyond-diagonal RIS architecture. We considered an RIS partitioned into multiple groups that are activated sequentially, which enables channel estimation and reliable signal detection while handling the constraint especially with large RIS size with the conventional and non-grouped RIS. In this system, we developed a group-wise maximum likelihood channel estimation approach and subsequently optimized the RIS phase shifts to maximize the effective received signal power. We presented both GLRT and ED based spectrum sensing methods by aggregating the TS across RIS groups. The
numerical results demonstrate that the proposed grouped RIS architecture achieves robust sensing performance even for small RIS sizes. In particular, the GLRT consistently performs better than ED across range of SNR values, while optimally configured RIS phase shifts provide apparent performance compared to random phase settings. The impact of correlated noise was also explored and the results confirm the robustness of the proposed system.
\bibliographystyle{IEEEtran} 
\bibliography{References}
\appendices
\section{ML Estimation of $\mathbf{h}_g$ and $\sigma_x^2$}
\label{appendix:ML_derivation}
\vspace{-2mm}
The log-likelihood function of the unknown parameters $\mathbf{h}_g$ and
$\sigma_{x}^2$ for the $g$-th RIS group, given the sample covariance matrix
$\hat{\mathbf{R}}_{\tilde{y},g}
= \tfrac{1}{T_g}\tilde{\mathbf{Y}}_g\tilde{\mathbf{Y}}_g^H$,
is given in \eqref{eq:logL_group} as
\begin{equation}
    \mathcal{L}_g(\mathbf{h}_g,\sigma_{x}^2)
    = -T_g \log|\boldsymbol{\Sigma}_{\tilde{y},g}|
      - T_g\,\mathrm{tr}\left(
      \boldsymbol{\Sigma}_{\tilde{y},g}^{-1}
      \hat{\mathbf{R}}_{\tilde{y},g}
      \right),
    \label{eq:appendix_logL_g}
\end{equation}
where $\boldsymbol{\Sigma}_{\tilde{y},g}$ is given in \eqref{eq:SigmaW_group}.
Using the matrix determinant lemma and Woodbury identity, we obtain
\begin{align}
    |\boldsymbol{\Sigma}_{\tilde{y},g}|
    &= 1 + \sigma_x^2
       \mathbf{h}_g^H\mathbf{A}_g^H\mathbf{A}_g\mathbf{h}_g,
    \label{eq:appendix_detlemma_g} \\
    \boldsymbol{\Sigma}_{\tilde{y},g}^{-1}
    &= \mathbf{I}
    - \frac{
      \sigma_x^2
      \mathbf{A}_g\mathbf{h}_g\mathbf{h}_g^H\mathbf{A}_g^H
      }{
      1 + \sigma_x^2
    \mathbf{h}_g^H\mathbf{A}_g^H\mathbf{A}_g\mathbf{h}_g
      }.
    \label{eq:appendix_woodbury_g}
\end{align}
Substituting \eqref{eq:appendix_detlemma_g}–\eqref{eq:appendix_woodbury_g}
into \eqref{eq:appendix_logL_g} yields
\begin{align}
    \mathcal{L}_g(\mathbf{h}_g,& \sigma_{x}^2)
    = -T_g \log\left(
       1 + \sigma_{x}^2
       \mathbf{h}_g^H\mathbf{A}_g^H\mathbf{A}_g\mathbf{h}_g
       \right)
       - T_g\,\mathrm{tr}(\hat{\mathbf{R}}_{\tilde{y},g})
       \nonumber\\
    &\quad
    + \frac{T_g\sigma_{x}^2}{
      1 + \sigma_{x}^2
      \mathbf{h}_g^H\mathbf{A}_g^H\mathbf{A}_g\mathbf{h}_g
      }
      \mathbf{h}_g^H
      \mathbf{A}_g^H
      \hat{\mathbf{R}}_{\tilde{y},g}
      \mathbf{A}_g
      \mathbf{h}_g.
    \label{eq:appendix_logL_final_g}
\end{align}
Now, we first obtain ML estimate of $\sigma_{x}^2$ for a given $\mathbf{h}_{g}$. Differentiating log-likelihood function, setting it equal to zero and performing some simplification yields the closed-form estimate of $\sigma_{x,g}^2$ as
\begin{equation}
    \hat{\sigma}_{x}^2(\mathbf{h}_g)
    =
    \frac{
    \mathbf{h}_g^H\mathbf{A}_g^H
    \hat{\mathbf{R}}_{\tilde{y},g}
    \mathbf{A}_g\mathbf{h}_g
    -
    \mathbf{h}_g^H\mathbf{A}_g^H\mathbf{A}_g\mathbf{h}_g
    }{
    \left(
    \mathbf{h}_g^H\mathbf{A}_g^H\mathbf{A}_g\mathbf{h}_g
    \right)^2
    }.
    \label{eq:appendix_sigma_est_g}
\end{equation}
Substituting \eqref{eq:appendix_sigma_est_g} into
\eqref{eq:appendix_logL_final_g} and ignoring constant terms gives
   $$ \mathcal{L}_g(\mathbf{h}_g,\hat{\sigma}_{x}^2)
    = T_g\left(
    \rho_g(\mathbf{h}_g) - \ln\rho_g(\mathbf{h}_g)
    \right),$$
where $
\rho_g(\mathbf{h}_g)
=
\frac{
\mathbf{h}_g^H\mathbf{A}_g^H
\hat{\mathbf{R}}_{\tilde{y},g}
\mathbf{A}_g\mathbf{h}_g
}{
\mathbf{h}_g^H\mathbf{A}_g^H\mathbf{A}_g\mathbf{h}_g
}.
$ Since the above expression is monotonically increasing in
$\rho_g(\mathbf{h}_g)$ for $\rho_g \ge 1$, maximizing the log-likelihood is equivalent to maximizing $\rho_g(\mathbf{h}_g)$. 
Thus, the ML estimate of $\mathbf{h}_g$ is given by
\begin{equation}
    \hat{\mathbf{h}}_g
    =
    \arg\max_{\mathbf{h}_g}
    \rho_g(\mathbf{h}_g).
\end{equation}
This is a generalized eigenvalue problem such that the solution satisfy
$\mathbf{A}_g^H
    \hat{\mathbf{R}}_{\tilde{y},g}
    \mathbf{A}_g
    \hat{\mathbf{h}}_g
    =
    \lambda_{\max,g}
    \mathbf{A}_g^H\mathbf{A}_g
    \hat{\mathbf{h}}_g.$
\section{Derivation of the GLRT Statistic}
\label{appendix:GLRT_derivation}
The covariance matrices of the received signal
$\tilde{\mathbf{y}}_{t}$ under the hypotheses in \eqref{eq:hypo_re_g}
are given by
\begin{align}
    \boldsymbol{\Sigma}_{0}
    &= \mathbb{E}[\tilde{\mathbf{y}}_{t}\tilde{\mathbf{y}}_{t}^H|\mathcal{H}_0]
    = \mathbf{I}, \\
    \boldsymbol{\Sigma}_{1}
    &= \mathbb{E}[\tilde{\mathbf{y}}_{t}\tilde{\mathbf{y}}_{t}^H|\mathcal{H}_1]
    = \mathbf{I}
    + \sigma_x^2
      \mathbf{A}^\star
      \mathbf{h}\mathbf{h}^H
      \mathbf{A}^{\star H}.
    \label{eq:appendix_glrt_cov_g}
\end{align}
The log-likelihood ratio (LLR) 
for given $\mathbf{h}$ and $\sigma_x^2$ becomes
\begin{align*}
    {\rm LLR}(\mathbf{h},\sigma_x^2)
    =
    &T\left(\,\mathrm{tr}
    (\hat{\mathbf{R}}_{\tilde{y}}) -\ln|\boldsymbol{\Sigma}_{1}|
    -\,\mathrm{tr}\left(
    \boldsymbol{\Sigma}_{1}^{-1}
    \hat{\mathbf{R}}_{\tilde{y}}
    \right) \right).
\end{align*}
The GLRT statistic is obtained by maximizing ${\rm LLR}$ with respect to
$(\mathbf{h},\sigma_x^2)$. Let
$
\mathbf{C} = \mathbf{A}^{\star H}\mathbf{A}^\star$
and
$
\mathbf{B} = \mathbf{A}^{\star H}
\hat{\mathbf{R}}_{\tilde{y}}
\mathbf{A}^\star.
$
Using the matrix determinant lemma and Woodbury identity,
$|\boldsymbol{\Sigma}_{1}|$ and $\boldsymbol{\Sigma}_{1}^{-1}$
can be simplified so that the LLR becomes
\begin{align}
    {\rm LLR}(\mathbf{h},\sigma_x^2)
    &= -T\ln\left(
    1+\sigma_x^2\mathbf{h}^H\mathbf{C}\mathbf{h}
    \right) \nonumber\\
    &\quad
    +T\,
    \frac{\sigma_x^2}{
    1+\sigma_x^2\mathbf{h}^H\mathbf{C}\mathbf{h}
    }
    \mathbf{h}^H\mathbf{B}\mathbf{h}.
    \label{eq:appendix_glrt_llr_expanded_g}
\end{align}
Following steps similar to Appendix~\ref{appendix:ML_derivation},
the above expression is maximized with respect to $\sigma_x^2$
for a given $\mathbf{h}$, giving
    $\hat{\sigma}_x^2(\mathbf{h})
    =
    \frac{
    \mathbf{h}^H\mathbf{B}\mathbf{h}
    -
    \mathbf{h}^H\mathbf{C}\mathbf{h}
    }{
    (\mathbf{h}^H\mathbf{C}\mathbf{h})^2
    }.$
Further, as discussed in Section~\ref{sec:estimation}, imposing the normalization
constraint $\mathbf{h}^H\mathbf{C}\mathbf{h}=1$ to remove  scale
ambiguity gives
$
    \hat{\sigma}_x^2(\mathbf{h})
    =
    \max\{\mathbf{h}^H\mathbf{B}\mathbf{h} - 1,\,0\}.
$
Substituting this into
\eqref{eq:appendix_glrt_llr_expanded_g}, we get
\begin{equation}
    {\rm LLR}(\mathbf{h})
    =
    T\left(
    \mathbf{h}^H\mathbf{B}\mathbf{h}
    - 1
    - \ln(\mathbf{h}^H\mathbf{B}\mathbf{h})
    \right),
    \label{eq:appendix_glrt_llr_profiled_g}
\end{equation}
which now needs to be maximized with respect to $\mathbf{h}$.
Maximizing \eqref{eq:appendix_glrt_llr_profiled_g}
subject to $\mathbf{h}^H\mathbf{C}\mathbf{h}=1$
leads to the generalized eigenvalue problem
    $\mathbf{B}\hat{\mathbf{h}}
    =
    \lambda_{\max}\mathbf{C}\hat{\mathbf{h}},$
where $\lambda_{\max}$ is the dominant generalized eigenvalue
of the matrix pair $(\mathbf{B},\mathbf{C})$.
Thus,  the GLRT statistic as
\begin{equation}
    {\rm TS}_{\mathrm{GLRT}}
    =
    T\left(
    \lambda_{\max}
    - 1
    - \ln \lambda_{\max}
    \right).
\end{equation}
Finally, the overall GLRT statistic can be obtained  as given in \eqref{eq:GLRT_statistic_g}.

\end{document}